\documentclass{Interspeech}

\interspeechcameraready

\title{SemAlignVC: Enhancing zero-shot timbre conversion using semantic alignment
\thanks{*Work done during internship at GenAI@Meta.}
}

\author[affiliation={1}]{Shivam}{Mehta*}
\author[affiliation={2}]{Yingru}{Liu}
\author[affiliation={2}]{Zhenyu}{Tang}
\author[affiliation={2}]{Kainan}{Peng}
\author[affiliation={2}]{Vimal}{Manohar}
\author[affiliation={2}]{Shun}{Zhang}
\author[affiliation={2}]{Mike}{Seltzer}
\author[affiliation={2}]{Qing}{He}
\author[affiliation={2}]{Mingbo}{Ma}

\affiliation{Division of Speech Music and Hearing}{KTH Royal Institute of Technology}{Sweden}
\affiliation{GenAI}{Meta}{USA}
\email{smehta@kth.se, \{yingrliu, zyt, kainanp, vimalmanohar, shunz, mikeseltzer, qinghe, mingboma\}@meta.com}

\newcommand{\Tau}{\tau}

\usepackage{comment}

\usepackage[
    colorlinks=true,
    linkcolor=blue,        
    citecolor=blue,        
    urlcolor=magenta       
]{hyperref}

\usepackage{cite}
\usepackage{graphicx}
\def\BibTeX{{\rm B\kern-.05em{\sc i\kern-.025em b}\kern-.08em
    T\kern-.1667em\lower.7ex\hbox{E}\kern-.125emX}}

\usepackage{bm} 
\usepackage{subcaption}  

\usepackage{booktabs}

\usepackage{enumitem} 
\setlist{nolistsep} 

\usepackage{multirow} 
\usepackage{rotating} 
\usepackage{microtype}
\begin{document}

\maketitle

\begin{abstract}
    
Zero-shot voice conversion (VC) synthesizes speech in a target speaker’s voice while preserving linguistic and paralinguistic content. However, timbre leakage—where source speaker traits persist—remains a challenge, especially in neural codec and LLM-based VC, where quantized representations entangle speaker identity with content. We introduce SemAlignVC, an architecture designed to prevent timbre leakage using SemAlign, a novel method that aligns text and audio representations to ensure speaker-independent semantic encoding. This disentangled representation conditions an autoregressive transformer for high-fidelity conversion without explicit speaker embeddings. Experiments show SemAlignVC significantly reduces timbre leakage, outperforming baselines in speaker timbre similarity, intelligibility, and naturalness, making it a robust, privacy-preserving, and generalizable VC solution.
\end{abstract}

\section{Introduction}


Voice conversion (VC) replicates speaker identity while preserving the semantics and paralinguistic features of speech \cite{berrak2021_vc_overview}. This identity includes accent \cite{zhao18b_l2_arctic}, style, emotion \cite{zhou2022_emotional_vc}, and timbre. Zero-shot VC, or any-to-any VC, extends this capability to unseen speakers, posing challenges due to the absence of paired source-target utterances. Our work focuses on timbre conversion within neural audio codecs \cite{zeghidour2021soundstream, defossez2023_encodec, zhang2023speechtokenizer, ju2024naturalspeech} and Large Language Models (LLMs) \cite{yang2023uniaudio, wang2023lmvc, du2024cosyvoice2}. The key challenge is preventing timbre leakage, where residual source speaker traits persist in the output while ensuring linguistic and paralinguistic fidelity. This issue stems from the entangled quantization of timbre, content, and style in audio tokens.


Modern zero-shot VC methods aim to disentangle semantic content and timbre using information bottlenecks \cite{qian2019autovc, ju2024naturalspeech, chen2023streaming, li2023freevc, qian20unsupervised, baas2023knnvc}. Some extract semantic features via phone-posteriorgrams (PPGs) \cite{kovela2023any, sun2016phonetic} or HuBERT tokens \cite{hsu2021hubert, yang2024streamvc, wang2023lmvc, huang2023make, yang2023uniaudio, guo2023quickvc, van2022comparison}, while speaker identity is modeled through speaker verification embeddings \cite{chen2022wavlm, desplanques2020ecapa}. Despite achieving high-quality speech, these methods suffer from timbre leakage, where residual source speaker traits persist due to the entanglement of linguistic and speaker cues in tokenized speech. Additionally, models relying on speaker embeddings raise privacy concerns, as these embeddings encode identifiable speaker traits, complicating anonymization while often failing to generalize in any-to-any speaker setting. This issue is even more pronounced in neural codec-based LLMs, where quantization further blends speaker identity with linguistic content. A more robust approach is needed—one that effectively mitigates timbre leakage while ensuring a clean separation of semantics and speaker characteristics.


To mitigate timbre leakage in zero-shot VC for neural codec-based LLMs while ensuring speaker privacy, we introduce SemAlignVC. This approach leverages SemAlign, a novel text-to-audio alignment strategy that filters timbre information from audio representations, enabling high-quality VC. Our key contributions are:
\begin{itemize}
    \item We show that widely used speech representations (e.g., HuBERT, EnCodec) inherently encode speaker information, leading to timbre leakage in zero-shot VC.
    \item We introduce SemAlign, a method that enforces semantic-text alignment to remove speaker identity cues.
    \item We propose SemAlignVC, a zero-shot VC model that eliminates the need for explicit speaker embeddings, enhancing privacy and zero-shot conversion.
    \item Objective and subjective evaluations confirm that SemAlignVC outperforms existing zero-shot VC models for timbre conversion while maintaining high-quality speech.
\end{itemize}
SemAlignVC effectively removes timbre leakage while preserving semantic integrity, making it well-suited for privacy-sensitive applications. Audio samples are available at demo page \href{https://shivammehta25.github.io/SemAlignVC}{https://shivammehta25.github.io/SemAlignVC}

\section{Background}

\subsection{Speech Quantization}

Speech signals are sampled at high rates, generating thousands of data points per second—posing challenges for LLM-based modeling. To manage this complexity, vector quantization (VQ) techniques \cite{van2017neural} reduce the data rate by tokenizing audio. Common methods include Residual Vector Quantization (RVQ) \cite{zeghidour2021soundstream, defossez2023_encodec} and K-Nearest Neighbor (KNN)-based approaches like HuBERT \cite{hsu2021hubert} and Best-RQ \cite{chiu2022_best_rq}. However, these audio codec models entangle linguistic, style, and timbre information, making it difficult to isolate specific characteristics. This entanglement complicates voice conversion, where timbre must be filtered while preserving content.

\subsection{Voice Conversion}

VC with parallel datasets is straightforward, but zero-shot VC on non-parallel data remains challenging. AutoVC \cite{qian2019autovc} pioneered this area using an autoencoder with a bottleneck, yet timbre leakage persisted. GAN-based approaches \cite{kaneko2019stargan, kaneko2019cyclegan} improved end-to-end VC but introduced complexity and training instability. Recent methods rely on pretrained bottlenecks via audio tokenization \cite{yang2023uniaudio, huang2023make, wang2021vqmivc, wang2023lmvc, zhu2023vec, yang2024streamvc} or PPGs \cite{kovela2023any, chen2023streaming}, but these still blend content and timbre, requiring additional masked training \cite{wang2023lmvc} or speaker embeddings from verification models. This reliance raises privacy concerns and often limits zero-shot generalization.

\begin{figure*}[ht]
    \centering
    \includegraphics[width=\textwidth]{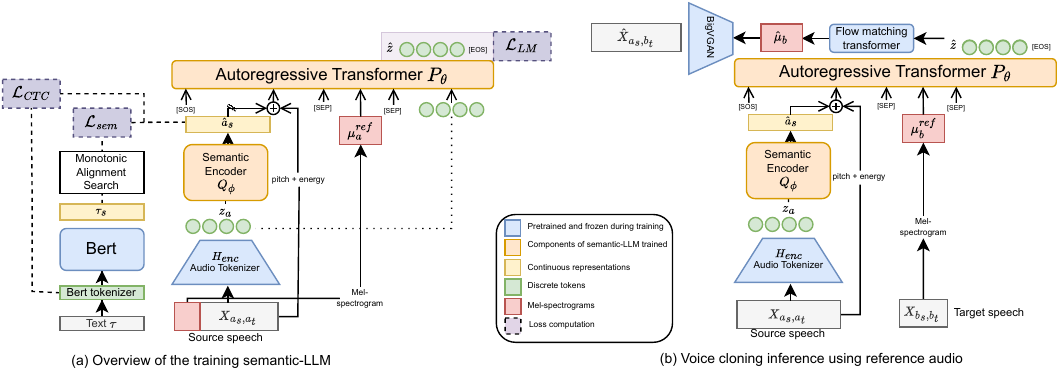}
    \caption{Schematic overview of training semantic LLM (left) and using the entire pipeline for voice conversion (right)}
    \label{fig: architecture}
    \vspace{-1em}
\end{figure*}

\section{Method}

In this section, we formalize the voice conversion framework for non-parallel datasets and introduce our approach, which consists of four individually trained components: audio tokenization, a semantic language model (semantic-LLM), an acoustic model, and a vocoder. Additionally, we describe two auxiliary objectives, the Connectionist Temporal Classification (CTC) loss \cite{graves2006connectionist} and the novel SemAlign methodology. These objectives help the semantic model learn a speaker-independent representation, crucial for disentangling semantic information from timbre effectively reducing timbre leakage. The overall architecture of SemAlignVC is depicted in Fig. \ref{fig: architecture}.
\subsection{Motivation}
Let $X_{a_s, a_t}$ represent an input utterance from speaker $A$, where $a_s$ and $a_t$ denote the semantic (linguistic and paralinguistic) component, and timbre component. For brevity, we focus on these components only as they are central to our work. The objective is to train a geneative model $P_\theta$ that reconstructs the utterance as $\hat{X}_{a_s, a_t} = P_\theta(X_{a_s, a_t})$. Critically, the model must disentangle $a_s$ and $a_t$ such that, during inference, it can combine the semantic content of one speaker with the timbre of another. For example, given an utterance $X_{b_s, b_t}$ from speaker $B$, the model should synthesize $\hat{X}_{a_s, b_t} = P_\theta(X_{a_s, a_t}, X_{b_s, b_t})$. However, since $P_\theta$ is not exposed to this combination during training, disentangling $a_s$ and $a_t$ becomes critical for generalization.

\subsection{Model Architecture}
For \textbf{audio tokenization}, we employ BEST-RQ-based \cite{chiu2022_best_rq} VQ-VAE to quantize continuous speech signals into tokens, effectively downsampling them. The encoder of our quantizer generates token IDs $z_a = H_{Enc}(X_{a_s, a_t})$, where $z_a$ encapsulates both timbre and semantic components, further entangling them. The tokens $z_a$ are extracted from a pretrained audio tokenizer and passed to the \textbf{semantic-LLM}, which is responsible for content modeling and timbre disentanglement. During training, a semantic encoder $Q_\phi$ aims to retrieve semantic and linguistic information from $z_a$, such that $\hat{a}_s = Q_\phi(z_a)$. We also extract pitch ($f_{0}$) and energy ($e$) from the input audio to preserve its paralinguistic information and use utterance-level mean normalization to remove speaker-related information \cite{lameris2023prosody}. These extracted features are fed to the LLM to learn the distribution $P_\theta(\hat{z}_a | \hat{a}_s, f_0, e)$, where ideally $\hat{z}_a \approx z$, with $z$ representing the speaker-independent tokens that contain the semantic and linguistic information of the input audio.

To ensure that the semantic encoder $Q_\phi$ filters timbre, we introduce two auxiliary losses during training. Inspired by prior work \cite{kovela2023any, sun2016phonetic, anastassiou2024voiceshop}, we employ CTC Loss to encourage $Q_\phi$ to retain semantic information, similar to predicting PPGs, we tokenize $\tau$ using BERT Tokenizer\footnote{https://huggingface.co/google-bert/bert-base-uncased}. However, we found that CTC alone could not effectively disentangle timbre information. We suspect this is due to the nature of the CTC training inherently allowing blank tokens can retain unwanted speaker characteristics when modeled by a strong decoder. 
To explicitly remove such cues, we introduce \textbf{SemAlign}, which uses the text $\Tau$ of the utterance $X_{a_s, a_t}$ and pass it through a pretrained text-only semantic encoder (BERT in our case) to obtain $\tau_s = \operatorname{BERT}(\Tau)$ and align this with the output $Q_\phi$ using Monotonic Alignment Search (MAS) \cite{badlani2022one, mehta2022neural, mehta2023overflow}. Unlike CTC, which learns a phoneme-based representation that may still contain speaker-specific spectral features, SemAlign enforces direct text-to-audio mapping, ensuring that no timbre-related cues remain. To match the channel dimensions of $\tau_s$ and $\hat{a}_s$, we extend $\tau_s$ by repeating it. This temporal alignment upsamples $\tau_s$ to match the temporal dimension of $\hat{a}_s$, and we minimize the mean square error ($\mathcal{L}_{sem}$) between the upsampled $\tau_s$ and $\hat{a}_s$. This process forces the semantic encoder to learn only the semantic component while effectively removing any speaker information. Similarly to \cite{yang2024streamvc}, we restrict gradient propagation from the LLM decoder to the content encoder, ensuring that speaker information does not inadvertently leak into the learned representations. During training, we segment each utterance into two parts: a primary input audio $X_{a_s, a_t}$ and a smaller excerpt, approximately 25\% of its total length. The mel spectrogram of this smaller segment, denoted as $\mu_{a}^{ref}$, serves as the timbre reference. At inference time, this serves as the timbre of the target speaker. Finally, the LLM is trained with the prompt $\left[ [\text{SOS}], \hat{a}_s, [\text{SEP}], \mu_{a}^{ref}, [\text{SEP}], z_a, [\text{EOS}] \right]$, where the cross entropy loss ($\mathcal{L}_{LM}$) is computed only for $z_a$ and $[\text{EOS}]$.

Separately, we train an \textbf{acoustic model} to transform audio tokens ($z_a$) into mel spectrograms ($\mu_a$). We adopt a conditional flow matching-based transformer \cite{lipman2023flow, mehta2024matcha, le2024voicebox} for this purpose. Similar to VoiceBox \cite{le2024voicebox}, we employ a temporal span masking strategy during training. We extract $z_a$ from a pretrained audio tokenizer and learn to model the conditional distribution $p(\mu_a^{\text{mask}} | \mu_a^{\text{ctx}}, z_a)$, where $\mu_a^{\text{mask}}$ is the target for reconstruction and $\mu_a^{\text{ctx}}$ serves as the timbre reference. However, if $z_a$ contains speaker timbre information, the flow matching model, despite its robustness, may not generalize well when inferring for another speaker's tokens, $z_b$. Here, SemAlign is crucial, as it ensures that the predicted $\hat{z}_a \approx \hat{z}$ is speaker-independent, allowing the acoustic model to rely solely on the timbre from $\mu_a^{\text{ctx}}$. During VC inference, we pass $\mu_{b}^{\text{ref}}$ to both the semantic-LLM and the context of the acoustic model, resulting in $\hat{\mu}_b$ as the output. We use BigVGAN as the \textbf{vocoder} to convert $\hat{\mu}_b$ into the final waveform $\hat{X}_{a_s, b_t}$.

\section{Experiment}

\subsection{Speaker Classification Using Acoustic Representations}
To motivate and demonstrate the effectiveness of SemAlign, we experimented using pre-trained acoustic representations, both discrete and continuous and trained a speaker classification head on top. This was done to evaluate the presence of speaker information. We utilized discrete token IDs from EnCodec \cite{defossez2023_encodec}, HuBERT base\footnote{https://huggingface.co/facebook/hubert-base-ls960} \cite{hsu2021hubert} (specifically layers 9 as used by UniAudio\cite{yang2023uniaudio}), and our audio tokenizer $H_{Enc}$ . These token IDs were passed through an embedding layer $\in \mathbb{R}^{d}$ where $d$ is the dimension of the hidden layer of LLM, followed by a classification head containing a single layer. Additionally, we used continuous representations from WavLM and our trained semantic encoder $Q_\phi$, which were directly fed into a classification head. We trained these models using the LibriHeavy dataset for 5 epochs on 8 H100 GPUs. The results are presented in Table \ref{table: speaker classification}.


\begin{table}
\centering
\caption{Test accuracy for speaker classification using various acoustic representations. Lower values indicate less speaker information}
\begin{tabular}{llc}
\toprule
\textbf{Model} & \textbf{Type of representation} & \textbf{Accuracy (\%)} \\
\midrule
Encodec                  &   Discrete    &    96.7 \\
HuBERT (layer 9)         &   Discrete    &    71.7\% \\
$\operatorname{Ours}_{\text{tok}}$         &   Discrete    &    82.05\% \\
\midrule
WavLMXVectors            &   Continous    &    88.4\% \\
$\operatorname{Ours}_{Q_\phi}$  & Continous &    \textbf{2.84\%} \\

\bottomrule
\end{tabular}
\label{table: speaker classification}
\vspace{-8pt}
\end{table}

\subsection{Voice conversion framework}
Our audio tokenizer follows the architecture outlined in \cite{chiu2022_best_rq} and is trained on a large internal dataset. It can be easily substituted with other tokenization mechanisms, as they all use VQ-VAE type quantization that blends content and speaker information.

We train English-only semantic model from scratch on 60k hours of ASR-transcribed English audiobooks, which is the same dataset used by \cite{le2024voicebox} using 16 nodes containing 8 H100s for 350k iterations. Our LLM uses the same architecture as LLaMa, but we shrink the hidden dimension size $d \in \mathbb{R}^{2048}$ and set the number of layers to $8$, resulting in overall 0.5B parameters. For semantic encoder $Q_\phi$ we use 4 layers of conformer with hidden dimension $d$, we use beta-binomial prior and alignment losses from \cite{badlani2022one} to improve the temporal alignment. We use AdamW optimizer with a peak learning rate of 1e-4 warmed up for 2k steps and decays after. For the acoustic model, the architecture is identical to the VoiceBox \cite{le2024voicebox}, similarly, we use a Midpoint ODE solver with 12 steps (NFE=24).

To demonstrate the effectiveness of SemAlign, we compared SemAlignVC against three baseline models: KNNVC \cite{baas2023knnvc}, HierSpeech++ \cite{lee2023hierspeech++}, and UniAudio \cite{yang2023uniaudio}. We utilized their official code repositories and pre-trained checkpoints available on GitHub. For subjective evaluation, we selected four sentences from ten unseen speakers in the VCTK test set. Since individual utterances were only 3–4 seconds long, they were often too simplistic to capture the full complexity of speaker timbre and the nuances of voice conversion. To address this, we concatenated two sentences to form a single long utterance, resulting in 20 input utterances. Reference utterances were constructed by randomly selecting six unseen speakers (three male and three female) from the VCTK test set, resulting in a total of 20x6=120 audio samples. To assess speaker similarity, we conducted a Similarity Mean Opinion Score (SMOS) evaluation for speaker similarity. Each participant evaluated 10 audio samples per model, and with 120 participants, this resulted in a total of 1,200 ratings per model. Trained participants were asked to rate the speaker similarity between the reference and converted audio using a 5-point Likert scale, where 5 indicated "Excellent" (the voices sound identical), and 1 indicated "Bad" (the voices sound completely different). The results of this subjective evaluation are reported in Table \ref{table: subjective evaluations}.

For objective evaluation, we selected 50 utterances from the unseen speakers of LibriHeavy test set, which contained more spontaneous and complex intonation patterns compared to VCTK. For reference, we used ten different utterances from different unseen speakers, generating a total of 50x10 utterances per model. We then assessed these samples using multiple metrics. For naturalness, we employed DNSMOS \cite{reddy2022dnsmos}, which evaluates speech quality (\textbf{SIG}), background noise (\textbf{BAK}), and overall quality (\textbf{OVRL}). Intelligibility was measured via Word Error Rate (\textbf{WER}), using transcriptions generated by Whisper Large v3\footnote{https://github.com/openai/whisper}. We also evaluate $f_0$ consistency using Pearson correlation (\textbf{FPC}). Most importantly, to evaluate speaker and timbre similarity, we computed cosine similarity scores between reference and converted audio using three widely adopted speaker embedding models: WavLMXVectors (\textbf{WavLM}) \cite{chen2022wavlm}, ECAPA-TDNN (\textbf{ECAPA}) \cite{desplanques2020ecapa}, and Resemblyzer\footnote{https://github.com/resemble-ai/Resemblyzer} (\textbf{Resemb}). The results of the objective evaluation are reported in Table \ref{table: objective evaluation}.

\begin{table}[t]
    \centering
    \caption{SMOS Subjective Evaluation Results with their 95\% confidence intervals.}
    \label{table: subjective evaluations}
    \begin{tabular}{l c}
        \toprule
        \textbf{Model} & \textbf{SMOS Score ($\uparrow$)} \\
        \midrule
        KNNVC         & 2.77 $\pm$ 0.09 \\
        HierSpeech++  & 3.16 $\pm$ 0.12 \\
        UniAudio      & 2.56 $\pm$ 0.10  \\
        SemAlignVC    & \textbf{3.29 $\pm$ 0.09} \\
        \bottomrule
    \end{tabular}
\end{table}

\begin{table*}[ht]
    \centering
    \caption{Objective evaluation results for baseline voice conversion models compared against SemAlignVC. Evaluated for naturalness, consistency to input, intelligibility and speaker similarity WER \% for input samples was 6.77}
    \label{table: objective evaluation}
    \begin{tabular}{lccc|c|c|ccc}
        \toprule
        \textbf{Model} & \multicolumn{3}{c}{\textbf{Naturalness}} & \multicolumn{1}{|c}{\textbf{Consistency}} & \multicolumn{1}{|c}{\textbf{Intelligibility}} & \multicolumn{3}{|c}{\textbf{Speaker Similarity}} \\
        \cmidrule(lr){2-4} \cmidrule(lr){5-5} \cmidrule(lr){6-6} \cmidrule(lr){7-9}
        & SIG($\uparrow$) & BAK($\uparrow$) & OVRL($\uparrow$) & FPC($\uparrow$) & WER~\%($\downarrow$) & WavLM($\uparrow$) & ECAPA($\uparrow$) & Resemb($\uparrow$) \\
        \midrule
        KNNVC & 3.48  & 3.67 & 3.04 & \textbf{0.632} & 18.85 & 0.92  & 0.78 & 0.80 \\
        HierSpeech++ & 3.62 & 3.95 & 3.29 & 0.614 & \textbf{8.24} & 0.90 & 0.77 & 0.85 \\
        UniAudio & 3.62 & 4.06 & 3.34 & 0.610 & 9.98 & 0.75 & 0.42 & 0.66 \\
        \midrule
        SemAlignVC & \textbf{3.63} & \textbf{4.13} & \textbf{3.38} & 0.622 & 12.31 & \textbf{0.95} & \textbf{0.82} & \textbf{0.89} \\
        \bottomrule
    \end{tabular}
\end{table*}

\section{Results and Discussions}

\begin{figure}[t]
    \centering
    \begin{subfigure}{\linewidth}
        \centering
        \includegraphics[width=0.9\linewidth]{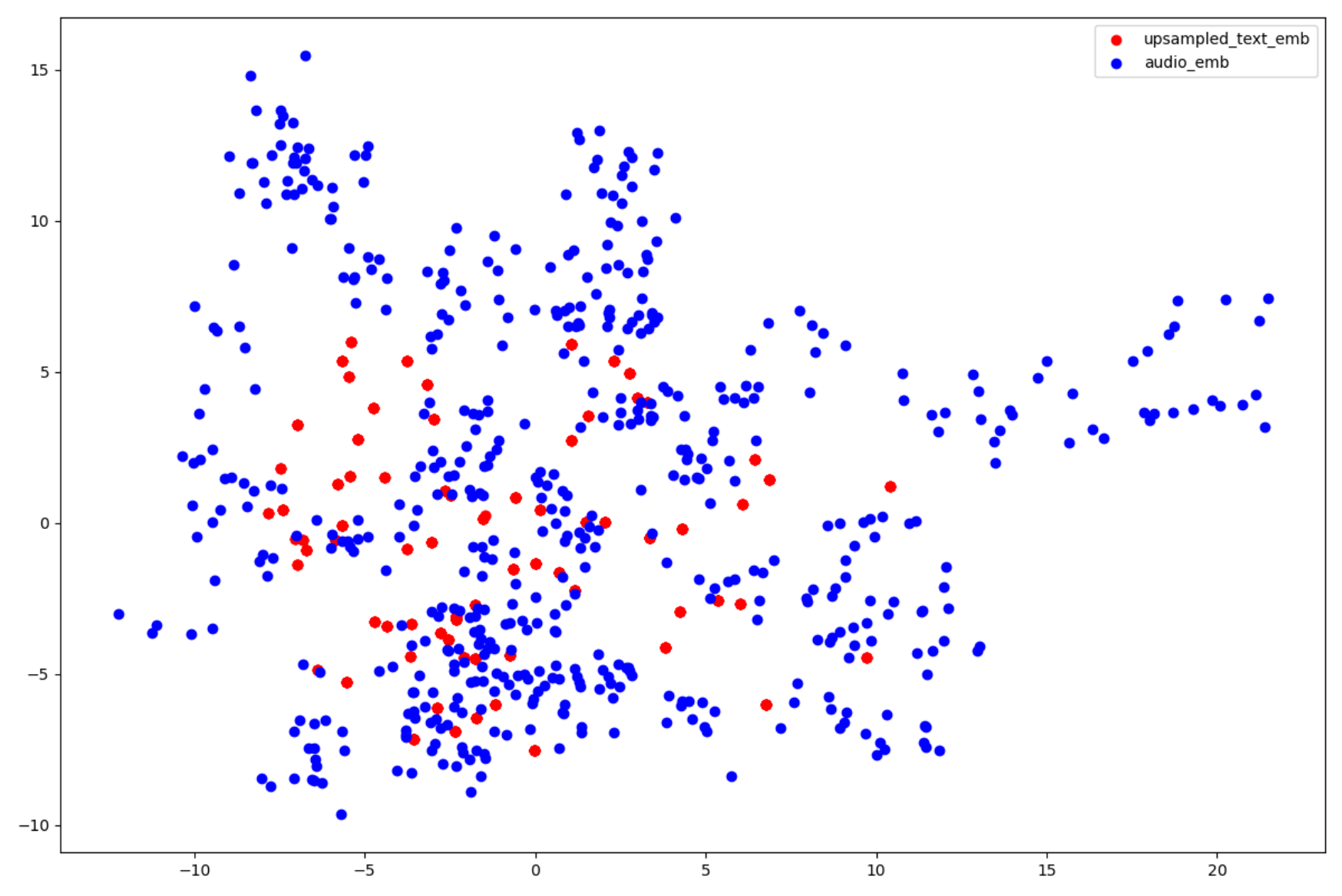}
        \label{fig: pca_1}
    \end{subfigure}
    

    \begin{subfigure}{\linewidth}
        \centering
        \includegraphics[width=0.9\linewidth]{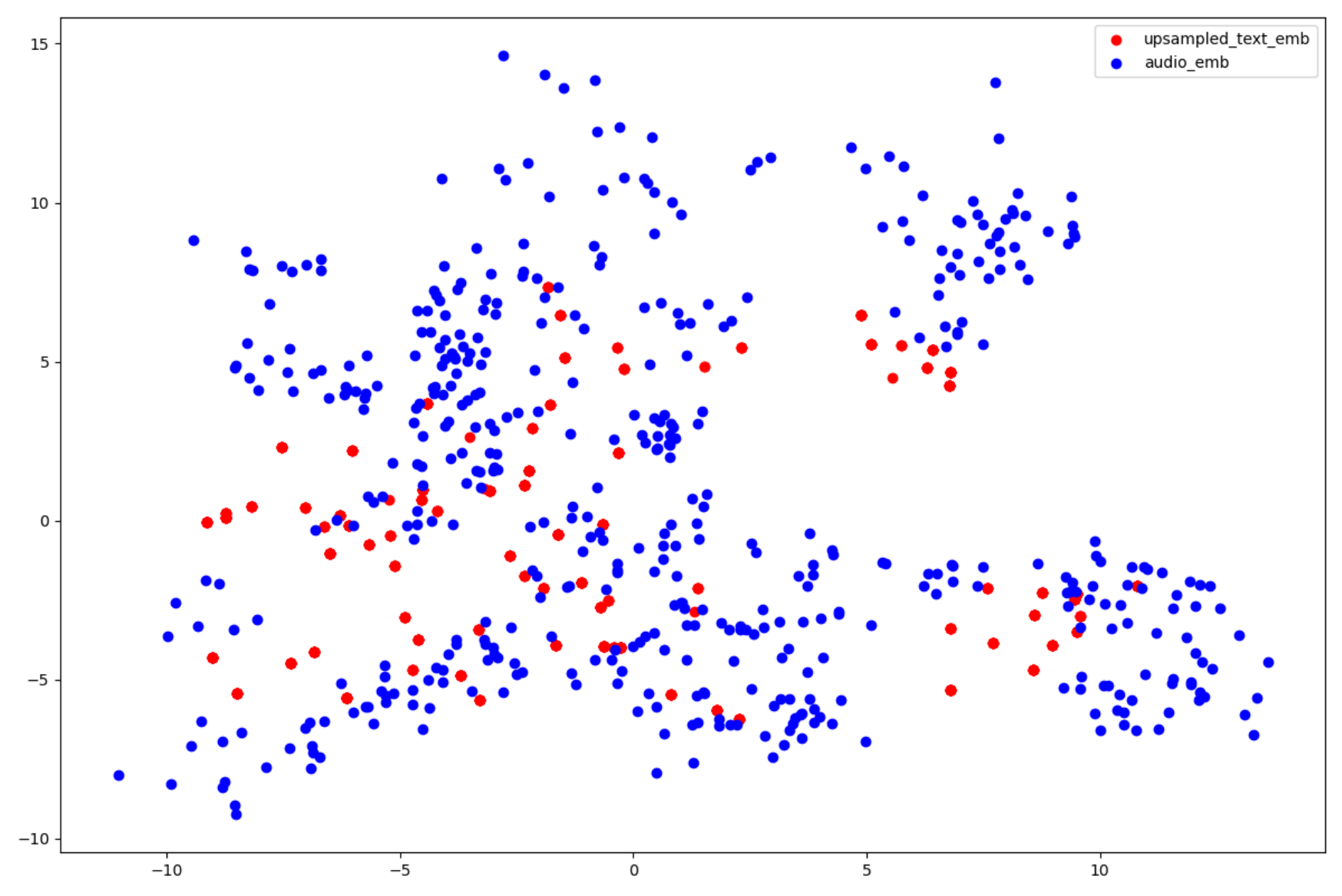}
        \label{fig: pca2}
    \end{subfigure}
    
    \caption{Principal Component Analysis (PCA) comparison of $Q_\phi$ (audio) depicted in blue and $\tau_s$ (text) embeddings depicted in red for two utterances.}
    \label{fig: PCA of audio embedding and text}
    \vspace{-1.5em}
\end{figure}

From Table \ref{table: speaker classification}, we derive two key observations. First, commonly used audio codecs and semantic encoders inherently retain a significant amount of speaker information, which can inadvertently lead to timbre leakage, thereby degrading the quality of voice conversion. This suggests that without proper disentanglement techniques, speaker identity can persist in representations intended to be speaker-independent. Second, our model, trained with SemAlign, demonstrates superior timbre removal capabilities, as evidenced by the low classification accuracy. This result indicates that the learned embeddings are nearly devoid of speaker-specific features, ensuring a more effective voice conversion process without unintended speaker timbre leakage. To further analyze the information captured by the semantic encoder $Q_\phi$ and compare it with the text-derived embedding $\tau_s$, we perform a Principal Component Analysis (PCA) on both representations. The results, visualized in Fig. \ref{fig: PCA of audio embedding and text}, reveal a strong alignment between the principal components of the audio embeddings and their textual counterparts. This alignment further reinforces the effectiveness of our approach in ensuring that the extracted features primarily encode semantic content while minimizing speaker-dependent characteristics.

For subjective evaluation in Table \ref{table: subjective evaluations}, longer sentences provided trained raters with more context, making subtle timbre differences more discernible and leading to more competitive similarity scores compared to short 3-second  utterances, thereby improving evaluation reliability. We observe that SemAlignVC achieves the highest naturalness, only comparable to HierSpeech++, while significantly outperforming the LLM-based baseline UniAudio. This demonstrates that SemAlign effectively mitigates timbre leakage.

Table \ref{table: objective evaluation} indicates that SemAlignVC consistently outperforms baseline models in several performance aspects. For \textbf{naturalness}, measured using DNSMOS scores, SemAlignVC achieves the highest overall rating (OVRL: 3.38), outperforming KNNVC (3.04), HierSpeech++ (3.29), and UniAudio (3.34). This improvement is attributed to our disentanglement process, which mitigates timbre leakage and preserves high speech quality. In terms of \textbf{consistency}, evaluated via fundamental frequency correlation (FPC), KNNVC achieves the highest score (0.632), with SemAlignVC following closely at 0.622. This suggests that while SemAlignVC ensures robust speaker-independent embeddings, slight variations in prosody consistency remain, which could be further optimized. For \textbf{intelligibility}, assessed using Word Error Rate (WER), SemAlignVC achieves a WER of 12.31\%, which is significantly lower than KNNVC (18.85\%) but slightly higher than HierSpeech++ (8.24\%). This suggests that while SemAlignVC effectively removes timbre information, there is room for improvement in preserving fine-grained linguistic details. We hypothesize that this discrepancy arises due to artifacts in the semantic encoding process. Specifically, when we trained a pure TTS model with BERT-based input representations, we observed occasional word substitutions with synonyms during generation. This artifact may be contributing to minor mispronunciations in our approach, which could be mitigated by refining text embeddings to focus on more linguistically grounded representations. Lastly, in \textbf{speaker similarity}, SemAlignVC achieves the highest scores across all three speaker similarity metrics: WavLM (0.95), ECAPA-TDNN (0.82), and Resemblyzer (0.89). This strongly suggests that the model effectively generalizes to mismatched audio input and speaker timbre, despite not encountering such combinations during training. These results demonstrate the effectiveness of SemAlignVC in removing timbre information from the source audio and generating timbre solely from the conditioned speaker input.

\section{Conclusion}
In this work, we introduced SemAlignVC, an architecture for zero-shot voice conversion that effectively disentangles semantic content from speaker timbre using SemAlign, a novel approach to filter timbre information from input audio and minimize timbre leakage for voice conversion. Our evaluation demonstrated that SemAlignVC outperforms baseline models in terms of naturalness and speaker similarity while achieving comparable intelligibility and minimizing timbre leakage without the use of any external speaker embedding, making it a high-quality, robust, and generalizable voice conversion method applicable even in privacy-sensitive environments. 

\bibliographystyle{IEEEtran}
\bibliography{main}

\end{document}